\documentclass[12pt]{iopart} 

\usepackage{amstext}

\usepackage[dvips]{graphicx}

\begin{document}

\DeclareGraphicsExtensions{.eps}
\graphicspath{{figures/}}

\title[NRG study of the Anderson-Holstein model]{Numerical Renormalization Group
  Study of the Anderson-Holstein Impurity Model.}
\author{A.C. Hewson and D. Meyer}
\address{Dept. of Mathematics, Imperial College, London SW7 2BZ.}

\begin{abstract}
We present numerical renormalization group (NRG) calculations for a
single-impurity Anderson model with a linear coupling to a local phonon
mode. We calculate dynamical response functions, spectral densities,
dynamic charge and spin susceptibilities. Being non-perturbative, the
NRG is applicable for all parameter regimes.
Our calculations cover both
weak and strong electron-phonon coupling for zero and finite
electron-electron interaction. We interpret the high- and low-energy
features and compare our results to atomic limit calculations and
perturbation theory. In certain restricted parameter regimes for strong
electron-phonon coupling, a soft phonon mode
develops inducing a very narrow resonance at the Fermi level.
\end{abstract}

\section{Introduction}

The effects of lattice vibrations on the electronic properties of metals
is a well studied branch of condensed matter physics. These effects include
specific heat enhancements, strong thermal contributions to the
electrical resistivity, and in some materials, superconductivity at low
temperatures.  There remain, however, some outstanding problems in the
field. For example, the extensive studies of small polaron formation,
based on the model from the pioneering work of Holstein~\cite{Hol59}, have
given a very precise understanding of systems with one or two electrons
present, but there are no comparable results for systems with a finite electron
density. There are also many strongly correlated electron systems, such
as heavy fermions or valence fluctuations systems, where there is known
to be strong coupling between the electrons and the lattice vibrations,
in addition to the strong inter-electron interactions. The interplay
of phonon and electron effects in these systems has not received much
attention. The reason for the lack of progress with these problems is
due to the fact there are no reliable calculational techniques in the strong
coupling regime, where standard perturbational methods break down. Some
non-perturbational techniques have been developed and successfully
applied to strongly correlated impurity models but in general these
cannot be extended to models of lattice systems. However, with the
introduction of dynamical mean field theory (DMFT)~\cite{GKKR96}, it has
proved possible to extend some of these non-perturbative techniques to a
wide class of lattice models. This approach exploits the fact that
certain infinite dimensional lattice models can be mapped onto effective
impurity models, together with a self-consistency condition
~\cite{GKKR96}.  The calculations for the effective impurity models can
be carried out using an appropriate non-perturbative method and iterated
until self-consistency is achieved. Though the method is
exact in the large dimensionality limit for an appropriately scaled
model~\cite{MV89}, the method can be applied to a much wider class of models and
there is evidence that it provides a very good first order approximation
for three dimensional systems. (It clearly breaks down for
one-dimensional systems but one-dimensional systems are a class
apart.)\par Not all techniques for solving strongly correlated impurity
models can be used for the effective impurity models generated through
the application of DMFT. The methods must be able to handle the
dynamical properties of the impurity models, which rules out techniques
such as the Bethe ansatz. The method which has been used most
extensively for this type of calculation is the quantum Monte Carlo
method (QMC). This approach, however, does have the drawback of not
being capable of being applied to the very low temperature regime. A
method which can be used at $T=0$ is the numerical renormalization group
approach (NRG). As developed originally by Wilson for the Kondo model
~\cite{Wil75,KWW80a,KWW80b} it was used only for the calculation of thermodynamic
quantities but the approach was extended later to the calculation of the
dynamic properties of impurity models~\cite{SSK89,CHZ94}, which is required to be able to
use the method in conjunction with DMFT for lattice models. The DMFT
approach together with the NRG has been applied to the half-filled
Hubbard model to study the metal-insulator transition at $T=0$~\cite{SK94,Bul99}, and
these calculations have recently been extended to finite
temperatures~\cite{BCV01}. Our goal is to extend this approach to include the effects of
coupling to local lattice modes, and address some of the outstanding
questions concerning the interplay of strong correlation and strong
electron-phonon interactions. The results should provide some insights
into the physics of systems such as heavy fermions, valence fluctuators,
manganites, and the cuprate high temperature superconductors.

In
this paper we perform NRG calculations solely for an impurity model with
a coupling to a local phonon mode. This is a necessary first step to the
eventual aim of incorporating the self-consistency constraint of the
DMFT and the application to models for lattice systems. We introduce the
impurity model in the next section.

\section{The Anderson-Holstein Impurity Model}
\label{sec:anders-holst-impur}

The model we study is a single impurity Anderson model~\cite{And61} with a
linear coupling to a local phonon mode, as in the the Holstein model
~\cite{Hol59}. We will refer to this model as the Anderson-Holstein (A-H)
model. The corresponding Hamiltonian for the model is
\begin{equation}
  \label{ah}
\eqalign{
    H=&\sum_{\sigma}\epsilon_f f^{\dagger}_\sigma f^{}_\sigma 
    +Uf^{\dagger}_\sigma f^{}_\sigma f^{\dagger}_{\bar\sigma} 
    f^{}_{\bar\sigma} + \lambda (b^{\dagger}+b)(\sum_{\sigma}f^{\dagger}_\sigma f^{}_\sigma -1)
    \\
    &+\sum_{{\bf k},\sigma}V_{\bf k}(f^{\dagger}_\sigma c^{}_{{\bf k}\sigma}
    + c^{\dagger}_{{\bf k}\sigma}f^{}_\sigma)+\sum_{{\bf k}\sigma}
    \epsilon_{\bf k}c^{\dagger}_{{\bf k}\sigma}c^{}_{{\bf k}\sigma}
    + \omega_0 b^{\dagger}b,
    }
\end{equation}
which describes an impurity d level $\epsilon_d$, hybridized with
conduction electrons of the host metal via a matrix element $V_k$, with
an interaction term $U$ between the electrons in the localized f (or d)
state, where $n_{f,\sigma}=f^{\dagger}_\sigma f^{}_{\sigma}$ is the
occupation number of this state which is linearly coupled to a local
oscillator of frequency $\omega_0$.

A very similar model which does
not have the interaction term $U$, but does not quite correspond to the
A-H model with $U=0$ as the electrons are spinless, has been studied
previously~\cite{HN79,HN80}. Both the model with spin and the spin-less
model can be solved exactly in the case of one electron only in the
system\cite{HN79}. As the interaction $U$ does not come into play in
this situation the results of both models
are essentially equivalent. However, there is no
exact solution for the many-electron case even for the spin-less model and $U=0$.

We will solve the model~(\ref{ah}) using the numerical renormalization group
method. We will briefly outline the general NRG scheme which is applicable to any
impurity model with local interactions and coupling to local bosonic or
fermionic fields~\cite{KWW80a,KWW80b}. 
The conduction
band of the model is logarithmically discretized and mapped
onto a semi-infinite linear chain coupled to the impurity at one end
such that the hopping matrix elements along the
chain fall off exponentially ($\sim \Lambda ^{-n/2}$).
The many-body states and energy levels are calculated iteratively starting from
the impurity adding an extra site of the chain in each step. After about 5-6
steps, the number of states has to be truncated to the 300-800 lowest ones
depending on the application in each step. This method has been generalized to
the calculation of dynamic as well as static properties~\cite{SSK89,CHZ94}.
In
applying this approach to the model with phonons (Eq.~(\ref{ah})), a truncation
of higher-energy steps is already necessary at the initial step.
The probability distribution function of finding
$n$ excited phonons in the system has a clear maximum around
$n=\bar{n}=v/\omega_0$, as can be read off
equation~(\ref{atlim}), and falls off rapidly for larger
values of $n$.
So it is sufficient to keep only
states with $n<n_{\rm cutoff}$ where we have chosen an initial $n_{\rm cutoff}=4\bar{n}$.
We have used the discretization parameter
$1.7<\Lambda<2.$, and kept at least 600 states in each iteration (see
Ref.~\cite{KWW80a,BCV01}). 

To calculate the spectral function on the
real axis, we need
the Fourier transform of
the double-time one-electron Green's function for the local d-electron.
We use this in the standard form,
\begin{equation}
  G_{\sigma}(\omega)=
  {1\over {\omega-\epsilon_{\rm f} +i\Delta }-\Sigma_{\sigma}(\omega)},
  \label{gf}
\end{equation} 
where we have taken the hybridization function 
$\Delta(\omega)=\pi\sum\sb {k}| V\sb {k}|\sp 2\delta(\omega
-\epsilon\sb{k})$ 
to be independent of $\omega$, corresponding to wide conduction
band with a flat density of states. The self-energy
$\Sigma_{\sigma}(\omega)$ will be calculated, as in reference~\cite{BHP98},
from the ratio of two Green's functions. Due to the additional interaction
term with the phonon there are two contributions so that
$\Sigma_{\sigma}(\omega)=\Sigma^U_{\sigma}(\omega)+\Sigma^{\lambda}_{\sigma}(\omega)$,
with
\begin{equation}
  \label{eq:sigma}
  \Sigma_\sigma^{U}(\omega)=U{F_\sigma(\omega)\over G_\sigma (\omega)},\quad
  \Sigma_\sigma^{\lambda}(\omega)=\lambda{M_\sigma(\omega)\over G_\sigma
    (\omega)},
\end{equation}
where $F_{\sigma}(\omega)$ and $M_{\sigma}(\omega)$ are the higher order Green's functions,
\begin{equation} 
  F_{\sigma}(\omega)=\langle\langle f^{}_\sigma f^{\dagger}_{\bar\sigma} 
f^{}_{\bar\sigma} ,f^{\dagger}_\sigma \rangle\rangle_{\omega} ,\quad
M_\sigma(\omega)= \langle\langle f^{}_\sigma (b^{\dagger}+b)
,f^{\dagger}_\sigma\rangle\rangle_{\omega}.
\end{equation} 
Details of the
derivation of this result from the equations of motion are given in the
Appendix A. In Appendix B, more details regarding the calculation of
$M_\sigma(\omega)$ are presented.

We will also calculate the spectral density of the Green's
function of the phonon mode, $D(\omega)=\langle\langle(b^{\dagger}+ b),(
b^{\dagger}+ b)\rangle\rangle_{\omega}$.  This can be expressed in terms
of the local charge susceptibility,
\begin{equation}
  \label{phononprop}
  D(\omega)=D^0(\omega)
  +\lambda^2 D^0(\omega)\langle\langle \hat O,\hat
  O\rangle\rangle_{\omega}D^0(\omega),
\end{equation}
where
\begin{equation}
  D^0(\omega)={2\omega_0\over \omega^2-\omega_0^2}
\end{equation}
is the non-interacting phonon propagator, and $\langle\langle \hat
O,\hat O\rangle\rangle_{\omega}$ with $\hat
O=\sum_{\sigma}f^{\dagger}_\sigma f^{}_\sigma -1$ is the local dynamic
charge susceptibility.

It will be useful in interpreting the
numerical results to compare them with results from perturbation theory
in the relevant regimes. We will set up the general perturbational
approach in this section and give specific results later. Starting from
expression for the partition function $Z$ in the form,
\begin{equation}
  Z/Z_0=\langle Te^{-\int_0^\beta H_{\rm int}(\tau)d\tau}\rangle_0,
  \label{pf}
\end{equation}
where $T$ is this usual time-ordering operator and $H_{\rm int}(\tau)$
is the interaction part of the Hamiltonian (including the hybridization
term) in the interaction representation with $\tau$ as the imaginary
time variable and $\beta=1/T$, and $\langle\,\rangle_0$ denotes a
thermal expectation value with respect to the non-interacting part of
the Hamiltonian. We can remove explicitly the phonon operators and
rewrite (\ref{pf}) as
\begin{equation}
  Z/Z_0=\langle Te^{-\int_0^\beta (H^F_{\rm int}(\tau)+H_{\rm ret}(\tau))d\tau}\rangle_0,
  \label{pf2}
\end{equation}
where $H^F_{\rm int}$ is the purely fermionic interaction part of the Hamiltonian (\ref{ah})
and $H_{\rm ret}$ is an additional retarded local fermionic interaction  due to the
elimination of the phonon terms, which is given by
\begin{equation}
  H_{\rm ret}(\tau)=\lambda^2\int_0^\tau \hat O(\tau)\hat
  O(\tau')D^0(\tau-\tau')d\tau',
  \label{pf3}
\end{equation} 
where $D^0(\tau)$ is the non-interacting phonon Green's function in the imaginary time form.
Details of this derivation are given in Appendix C. \par
It will be convenient to convert this expression to a path integral over fermionic Grassmann
variables and  integrate out the conduction
electrons. We can then  write (\ref{pf}) in the form, 
\begin{equation}
  Z=\int\prod_\sigma {\cal D}(\bar f_\sigma){\cal D}(
  f_\sigma)e^{-\int_0^{\beta} {\cal L}_{\rm eff}(\tau)d\tau},
\end{equation}
where
\begin{equation}
\eqalign{
  {\cal L}_{\rm eff}(\tau)=&\int_0^{\beta}d\tau'  \sum_{\sigma}
  \bar
  f_{\sigma}(\tau)[G_{\sigma}^{(0)}(\tau-\tau')]^{-1}f_{\sigma}(\tau')+U
  n_{\uparrow}(\tau)n_{\downarrow}(\tau)\\
  &+{\lambda^2\over 2}\int_0^{\beta}d\tau' D^0(\tau-\tau')(\sum_{\sigma}
  n_\sigma(\tau) -1)(\sum_{\sigma'} n_{\sigma'}(\tau') -1),
}
\end{equation}
where $\bar f_\sigma(\tau)$, $f_\sigma(\tau)$ are Grassmann variables,
$n_{\sigma}(\tau)= \bar f_{\sigma}(\tau)f^{}_{\sigma}(\tau)$, and
$G_{\sigma}^{(0)}(\tau)$ is the imaginary-time of the non-interacting
one electron Green's function for the localized f-electrons. By
introducing a Grassmann source field for the f-electrons one can convert
this partition function into a generating function for the interacting local Green's
function $G_{\sigma}(\tau)$ in the standard way~\cite{NObook}. This gives a convenient
way
of generating the perturbation theory and Feynman diagrams for this
Green's function.

In the limit $\omega_0\to\infty$ such that
$v=\lambda^2/\omega_0$ remains finite the effective interaction becomes
an instantaneous one equal to $-v(\sum_\sigma n_\sigma-1)^2$.  This
interaction term can be absorbed into the standard Anderson model with
the terms linear in $n_\sigma$ being absorbed by modifying
$\epsilon_f\to\epsilon_f+v$, and the remaining part by changing $U$ to
$U-2v$.
In this way of taking the limit $\omega_0\rightarrow
\infty$, the ratio $v/\omega_0\to 0$. But there is also a
high frequency regime of interest in which both $\omega_0$ and $v$ are
large such that $v/\omega_0\sim 1$.
For this parameter range it is
convenient to apply the canonical transformation $\hat U^{-1}H\hat U$
with $\hat U$ given by
\begin{equation}
  \hat U=e^{-{\lambda\over \omega_0}(b^{\dagger}-b) (\sum_\sigma
    n_{\sigma}-1) },
\end{equation}
which is a displaced oscillator transformation such that
\begin{equation}
  \fl\tilde b=\hat U^{-1}b\hat U=b-{\lambda\over\omega_0}(\sum_\sigma
  n_{\sigma}-1),\quad {\rm and}\quad
  \tilde f_\sigma=\hat U^{-1}f_\sigma\hat
  U=e^{{-{\lambda\over\omega_0}(b^{\dagger}-b)}}f_\sigma.
\end{equation}
\begin{equation}
  \eqalign{
    \fl H'=\hat U^{-1}H\hat U =&\sum_{\sigma}\bar\epsilon_f f^{\dagger}_\sigma f^{}_\sigma
    +\bar Uf^{\dagger}_\sigma f^{}_\sigma f^{\dagger}_{\bar\sigma} 
    f^{}_{\bar\sigma} \\
    &+\sum_{{\bf k},\sigma}V_{\bf
      k}\left(e^{{\lambda\over\omega_0}(b^{\dagger}-b)}f^{\dagger}_\sigma
      c^{}_{{\bf k}\sigma} +
      e^{-{\lambda\over\omega_0}(b^{\dagger}-b)}c^{\dagger}_{{\bf
          k}\sigma}f^{}_\sigma\right )+\sum_{{\bf k}\sigma} 
    \epsilon_{\bf k}c^{\dagger}_{{\bf k}\sigma}c^{}_{{\bf k}\sigma}
    + \omega_0 b^{\dagger}b,
    }
  \label{aht}
\end{equation} 
where $\bar\epsilon_f=\epsilon_f+v$ and $\bar U=U-2v$. The exponential
term in the phonon operators, which multiplies the hybridization term,
arises because a cloud of phonons may be created or absorbed when the
f-electron occupation changes, due to the coupling to the lattice. If we
take the expectation value of this Hamiltonian in the zero phonon state
$|0\rangle$ then, as $\langle
0|e^{\pm{\lambda/\omega_0}(b^{\dagger}-b)}|0\rangle=e^{-v/2\omega_0}$,
the 
effective resonance width becomes
$\bar\Delta=e^{-v/\omega_0}\Delta$. For $v/\omega_0\approx 2$ or 3, this
would amount to a significant narrowing of the resonance for the
non-interacting model. This exponential renormalization factor is a
common feature of strongly coupled electron-phonon problems but, as was
shown in earlier work of Hewson and Newns~\cite{HN79,HN80} for the spinless model, this
resonance narrowing does not in general occur in the strong coupling
regime for this model, but only in a certain parameter regime.

In the zero hybridization limit for the transformed Hamiltonian in
equation (\ref{aht}) the electrons and phonons become decoupled. However
in this limit the calculation of the Green's function in terms of the original
f-electron operators is not trivial, as it involves the unitary
transformation $\hat U$ but it can be evaluated exactly by using the
well known techniques for calculating the expectation values of
exponentials with a linear combinations of bose creation and
annihilation operators~\cite{Mahbook}.  The result for $T=0$ is
\begin{equation}
  \label{atlim}
  \eqalign{
    \fl G_\sigma(\omega)=  e^{-v/\omega_0} \sum_{n=0}^{n=\infty}
    \frac{(v/\omega_0)^n}{ n!} \bigg[
      &{\langle (1-n_{f,\sigma})(1-n_{f,-\sigma})\rangle \over \omega-\bar\epsilon_f-n\omega_0}
      +{\langle (1-n_{f,\sigma})n_{f,-\sigma}\rangle \over \omega-\bar\epsilon_f-\bar U-n\omega_0} \\
       &+{\langle n_{f,\sigma}(1-n_{f,-\sigma})\rangle \over \omega-\bar\epsilon_f+n\omega_0}
        +{\langle n_{f,\sigma}n_{f,-\sigma}\rangle \over \omega-\bar\epsilon_f-\bar U+n\omega_0}\bigg],
    }
  \end{equation}
where the expectation values, such as $\langle
(1-n_{f,\sigma})(1-n_{f,-\sigma})\rangle$, are calculated using
(\ref{aht}) in the 'atomic' limit $V_{k}=0$ for $T=0$.

In the weak coupling regime we can use perturbation theory based on the original form of 
the model given in equation (\ref{ah}). Using the finite temperature perturbation expansion
for the self-energy $\Sigma^{\lambda}_\sigma(\omega)$, carried out  to order $\lambda^2$,
and analytically continued to real frequency $\omega$,  we obtain
\begin{equation}
\eqalign{
    \Sigma_\sigma^{\lambda(2)}(\omega)=&{2\lambda^2\over\omega_0}(1-\sum_\sigma
n_{f,\sigma}) \\
&+\lambda^2\int
\rho_{f,\sigma}(\epsilon)\left\{{{f(\epsilon)+n_{\rm b}(\omega_0)}
\over{\omega-\epsilon+\omega_0}}+{{1-f(\epsilon)+n_{\rm b}(\omega_0)}
\over{\omega-\epsilon-\omega_0}}\right\}d\epsilon,
}
\label{sigl}
\end{equation}
where $n_{\rm b}(\epsilon)$ and $f(\epsilon)$ are the Bose-Einstein and
Fermi-Dirac distribution functions. The occupation value $n_{f,\sigma}$
of the f-state with spin $\sigma$, and the corresponding spectral
density $\rho_{f,\sigma}(\omega)$, can in principle include
self-energy contributions from the local interaction term $U$. These
quantities to zero order in $U$ are given by
\begin{equation}
n^0_{f,\sigma}={1\over 2}-{1\over \pi} {\rm
  tan}^{-1}\left({\epsilon_f\over\Delta}\right),\quad
\rho^0_{f,\sigma}(\epsilon)={\Delta/\pi\over{(\epsilon-\epsilon_f)^2+\Delta^2}}.
\label{nf0}
\end{equation}
Substituting  these expressions into~(\ref{sigl}) for the self-energy at $T=0$,
we find
\begin{equation} 
  \eqalign{
   {\rm Re} \Sigma_\sigma^{\lambda(2)}(\omega)= {4\lambda^2\over
    \omega_0\pi}{\rm tan}^{-1}\left({\epsilon_f\over \Delta}\right)+\\
 \fl  {\lambda^2\over\pi}{\rm Im}\left\{{\rm ln}\left({\epsilon_f+i\Delta\over
        |\omega+\omega_0|}\right){1
      \over{\omega-\epsilon_f+\omega_0-i\Delta}}-{\rm ln}\left({\epsilon_f+i\Delta\over
        |\omega-\omega_0|}\right){1
      \over{\omega-\epsilon_f-\omega_0-i\Delta}}\right\}
}
\end{equation}
and
\begin{equation} \fl
{\rm Im} \Sigma_\sigma^{\lambda(2)}(\omega)= -\pi\lambda \left(
  (1-\Theta(\omega+\omega_0))\frac{\Delta}{(\omega+\omega_0-\epsilon_f)^2+\Delta^2} + \Theta(\omega-\omega_0) 
  \frac{\Delta}{(\omega-\omega_0+\epsilon_f)^2 + \Delta^2}\right).
\end{equation}

The corresponding expression for $\Sigma_{\sigma}^U(\omega)$ to order $U^2$ is 
\begin{equation}
  \fl \Sigma^{U(2)}_\sigma(\omega)=Un^{(2)}_{f,-\sigma}+ U^2\int 
  \tilde\rho_{f,\sigma}(\epsilon)\tilde\rho_{f,-\sigma}(\epsilon')\tilde\rho_{f,-\sigma}(\epsilon'')
  {D(\epsilon,\epsilon',\epsilon'')\over
    {\omega-\epsilon+\epsilon'-\epsilon''}}d\epsilon
  d\epsilon'd\epsilon'',
  \label{sigu}
\end{equation}
where
\begin{equation}
  D(\epsilon,\epsilon',\epsilon'')=f(\epsilon')-f(\epsilon')
  f(\epsilon'')+f(\epsilon)f(\epsilon'')-f(\epsilon)f(\epsilon').
  \label{ff}
\end{equation}
The effects of phonon scattering could be included in the f-spectral density
$\rho_{f,\sigma}(\epsilon)$, and $n_{f,\sigma}$, in evaluating
(\ref{sigu}). If terms of order $\lambda^2$ were to be included they
would generate corrections of order $\lambda^2U^2$. These contributions
would be distinct from those of the same order  generated
by including terms of order $U^2$ in evaluating $\Sigma^\lambda(\omega)$
from equation (\ref{sigl}). To zero order in $\lambda$ we use
(\ref{nf0}) in evaluating (\ref{sigu}). 
These results can be used to compare with the NRG ones in the weak coupling regime.

There is a large parameter space for this model so we will present the
NRG results in four sections, corresponding to the symmetric and
asymmetric models with $U=0$, and then the same models with $U\ne 0$.

\section{Symmetric $U=0$ Model}
\label{sec:symmetric-u=0-model}

\begin{figure}[h]
  \begin{center}
    \includegraphics[width=0.75\textwidth]{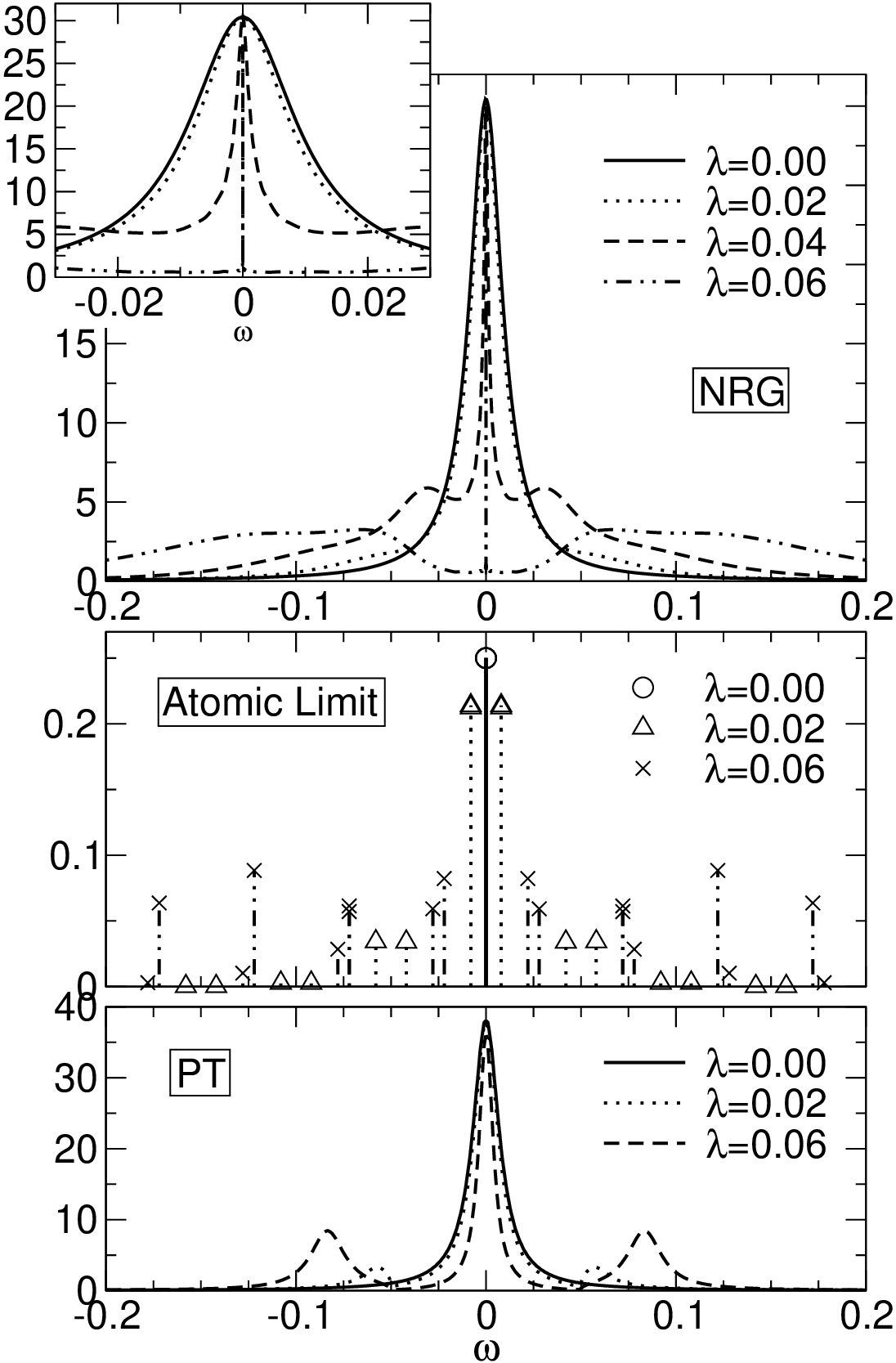}
    \caption{Spectral density
      $A(\omega)=-\frac{1}{\pi} \Im G_\sigma(\omega)$ for $U=0$,
      $\epsilon_f=0$, $V=0.1 \rightarrow \Delta\approx 0.016$ and
      $\omega_0=0.05$. The respective values of $\lambda$ are given in
      the graph. The upper panel shows NRG results, the two lower panels
      the atomic limit and the perturbation theory (PT) results. For the
      atomic limit calculation, the height of the peaks is a measure of
      the respective spectral weight of the excitation.}
    \label{fig:dos_u0sym}
  \end{center}
\end{figure}

\begin{figure}[ht]
  \begin{center}
    \includegraphics[width=0.5\textwidth]{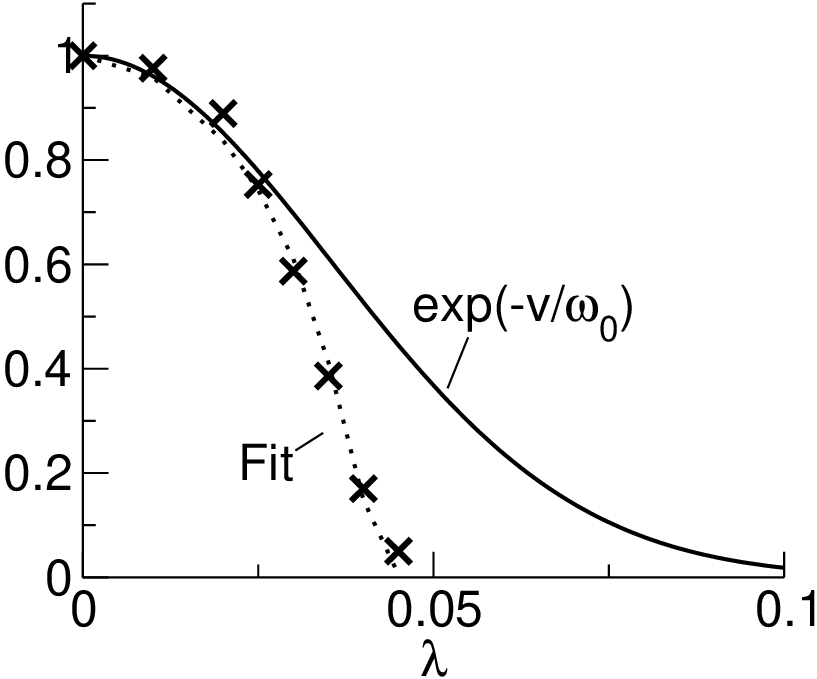}
    \caption{Width of central resonance ('zero-phonon peak') as function
      of $\lambda$ for the
      parameters of figure~\ref{fig:dos_u0sym} (NRG calculation). The solid and dashed lines
      are explained in the text.}
    \label{fig:peakwidth}
  \end{center}
\end{figure}

\begin{figure}[ht]
  \begin{center}
    \includegraphics[width=0.8\textwidth]{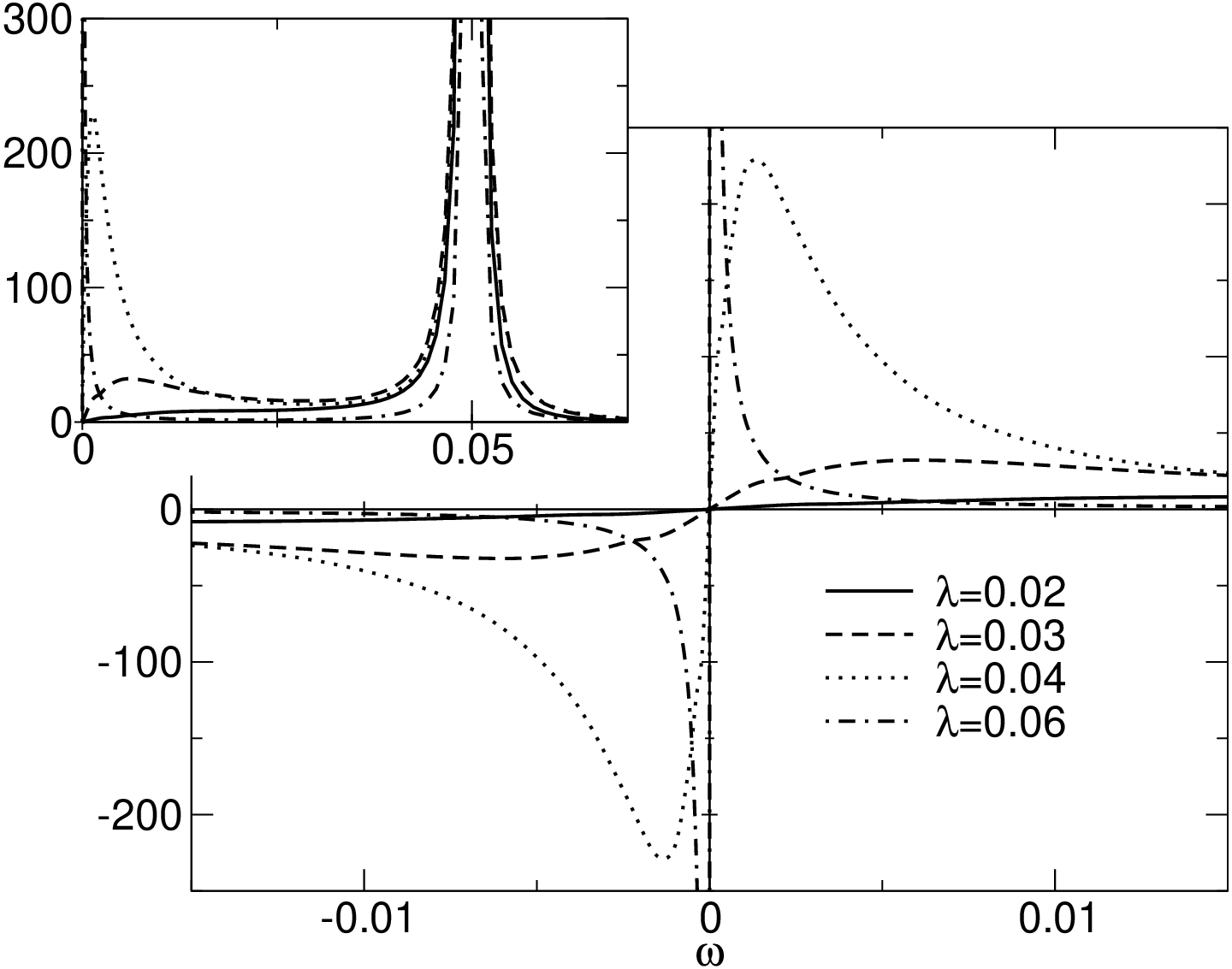}
    \caption{Imaginary part of the phonon propagator for the same
      parameters as used in figure~\ref{fig:dos_u0sym}. The inset shows
      the same quantity on a larger energy scale.}
    \label{fig:pp_u0sym}
  \end{center}
\end{figure}

\begin{figure}[ht]
  \begin{center}
    \includegraphics[width=0.6\textwidth]{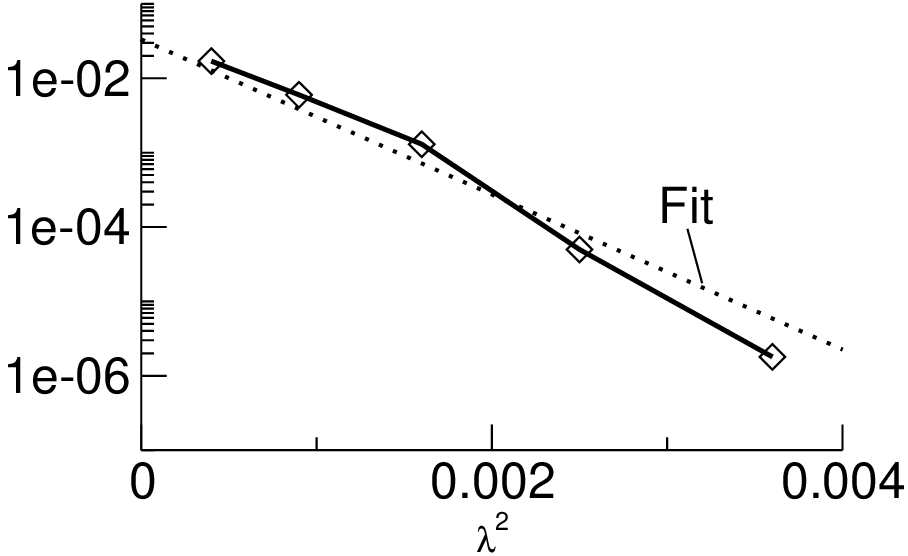}
    \caption{Position of the low-energy peak of the phonon propagator
      (see figure~\ref{fig:pp_u0sym}) plotted on a logarithmic scale vs. $\lambda^2$.}
    \label{fig:pp_peak}
  \end{center}
\end{figure}

In the upper panel of figure~\ref{fig:dos_u0sym} we present the NRG results for the spectral density of the
f-Green's function for various values of $\lambda$. The bare phonon
frequency has been taken
to be $\omega_0=0.05$ in this and all subsequent plots. The significant
features are the narrowing and virtual collapse of the central resonance
with increasing coupling strength $\lambda$ and the development and
outward shift of high energy peaks. This shift is proportional to
$\lambda^2$ and can be understood by looking at the corresponding
results in the atomic limit ($V=0$). The spectral weights and positions of the
atomic-limit excitation peaks of equation~(\ref{atlim}) for selected values of $\lambda$
are plotted in the middle panel of figure~\ref{fig:dos_u0sym}. The high-energy features in
the upper panel correspond to a broadening those seen in the atomic
limit. In this case $\bar{\epsilon_f} = v$ and $\bar{U}=-2v$, so the
effective interaction is an attractive one. The peaks below the Fermi
level correspond to removing an electron from a double-occupied site,
the peaks above to adding an electron to a previously unoccupied
site. The probability that the impurity is singly-occupied for finite
$\lambda$ vanishes exponentially with $\lambda^2$ (see
equation~(\ref{atlim})). The narrowing of the central peak, the
zero-phonon peak occurs very
rapidly with increase of $\lambda^2$.
On the basis of equation~(\ref{aht}), one might expect the narrowing to
follow the exponential form $\sim e^{-v/\omega_0}$. In figure~\ref{fig:peakwidth} we plot
the relative width of the central peak as function of $\lambda$. Only
for small values of $\lambda$ is it approximately proportional to
$\exp(-v/\omega_0)$. It decreases much more rapidly with higher coupling
strengths. This could be due to the effect of the softening of the
phonon mode.

The imaginary part of the phonon Green's function, displayed in
figure~\ref{fig:pp_u0sym}, shows clear evidence of the emergence of a soft
phonon mode already for intermediate coupling strength. 
The inset shows the same as the large figure, but over a larger
frequency range, displaying also the 'bare phonon' peaks at $\omega=\pm
0.05$. With increasing $\lambda$, another peak appears at lower
frequency. This soft phonon mode can be traced back to features in the
charge susceptibility.
Equation~(\ref{phononprop}) shows the relation between the phonon Green's
function and the dynamic charge susceptibility. The soft phonon mode
reflects the lower energy peak in the charge susceptibility.
Figure~\ref{fig:pp_peak} shows the position of the lower energy peak in phonon propagator.
It falls off
approximately with $\omega_\lambda=1/30 \exp(-\frac{(2.45*\lambda)^2}{\omega_0^2})$.
With this result, we also reach a good fit to the data of
figure~\ref{fig:peakwidth} if we assume the zero-phonon peak width to behave like
$\exp(-v(\frac{1}{\omega_0}+\frac{0.03}{\omega_\lambda})$
instead of $\exp(-v/\omega_0)$.

A comparison of our results with those of lowest-order perturbation
theory, equation~(\ref{sigl}), which is displayed in the lowest panel of figure~\ref{fig:dos_u0sym},
reveals the limitations of the perturbational
result in this parameter regime. We would not expect to get the
exponential narrowing of the zero-phonon peak from a weak-coupling
result, but the high-energy
features are also not well reproduced. Although there are high-energy
peaks, their positions are in poor agreement with the NRG results and
their shift to higher frequencies is proportional to $\lambda$ instead of
$\lambda^2$.

\section{Asymmetric $U=0$ Model}
\label{sec:asymmetric-u=0-model}

\begin{figure}[ht]
  \begin{center}
    \includegraphics[width=0.75\textwidth]{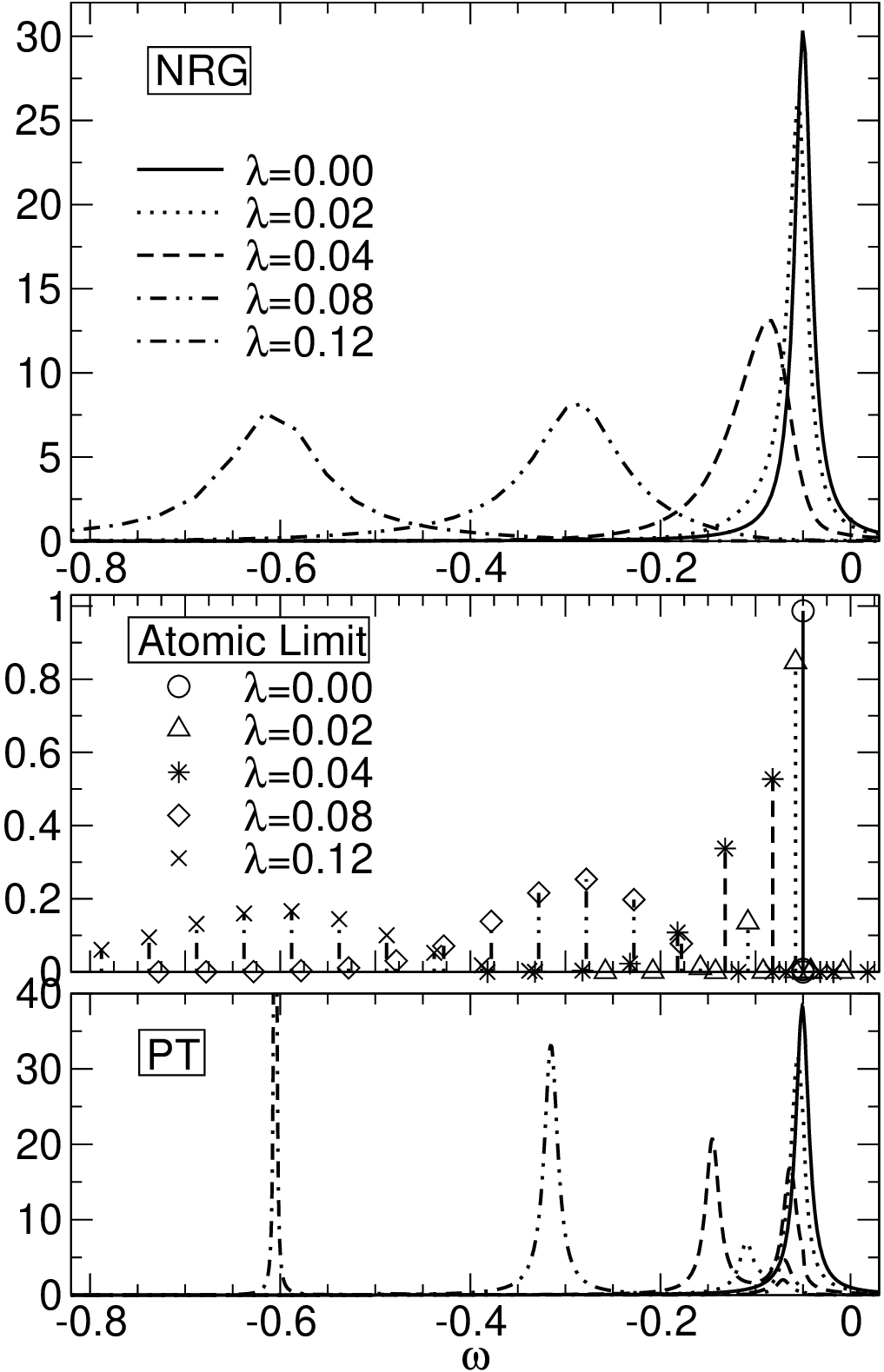}
    \caption{Same as figure~\ref{fig:dos_u0sym}, but with $\epsilon_f=-0.05$.}
    \label{fig:dos_u0asym}
  \end{center}
\end{figure}

\begin{figure}[ht]
  \begin{center}
    \includegraphics[width=0.6\textwidth]{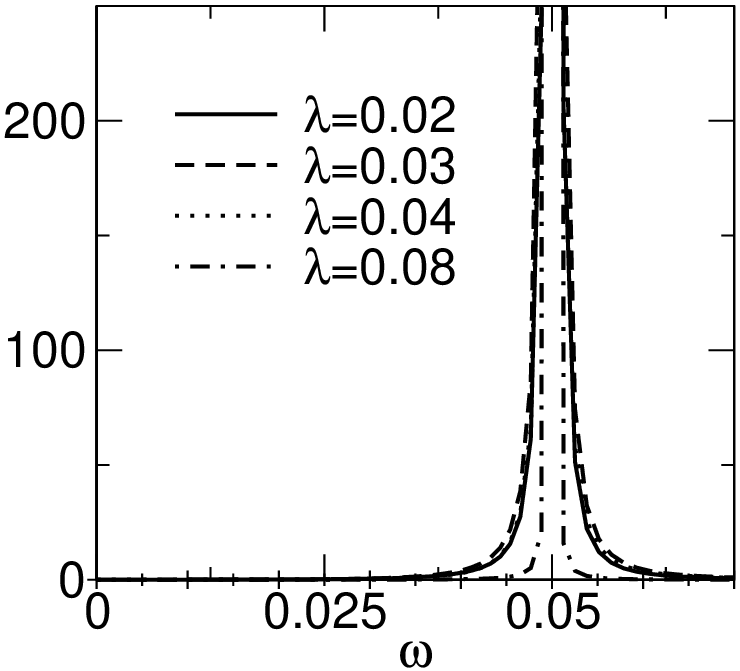}
    \caption{Imaginary part of the phonon propagator for the parameters
      of figure~\ref{fig:dos_u0asym}.}
    \label{fig:pp_u0asym}
  \end{center}
\end{figure}

As soon as we introduce an asymmetry into the model, we find a
qualitatively different picture. In figure~\ref{fig:dos_u0asym} we plot the
NRG, atomic limit and perturbation theory results for
the spectral density of the local Green's function for a range of
values of $\lambda$ similar to figure~\ref{fig:dos_u0sym}, but with an $f$-level located
at $\epsilon_f=-0.05$. In this spectrum there is a single peak
corresponding to an excitation from a doubly-occupied impurity site,
positioned at $\bar{\epsilon_f} +
\bar{U}-\bar{n}\omega_0=\epsilon_f-v-\bar{n}\omega_0$, where $\bar{n}=v/\omega_0$
is approximately the mean number of phonons excited.
The shift in the peak
to lower values reflects that seen in the atomic limit,
equation~(\ref{atlim}). Additionally, the peak broadens with increasing
$\lambda$ due to the higher probability of exciting multiple phonons on
removing the local electron. As the peak moves away from the Fermi
level, the charge fluctuations are suppressed. This is manifest in
the absence of the soft phonon mode, as can be seen in figure~\ref{fig:pp_u0asym}
showing the phonon propagator in this parameter range
(cf. equation~(\ref{phononprop})). 

In contrast to the symmetric case, the
perturbational result, shown in the lower panel of
figure~\ref{fig:dos_u0asym}, does give quite good agreement of the peak
positions, the increase of broadening, however, is absent.

\section{Symmetric $U\ne 0$ Model}
\label{sec:symmetric-une-0}

\begin{figure}[h]
  \begin{center}
    \includegraphics[width=0.75\textwidth]{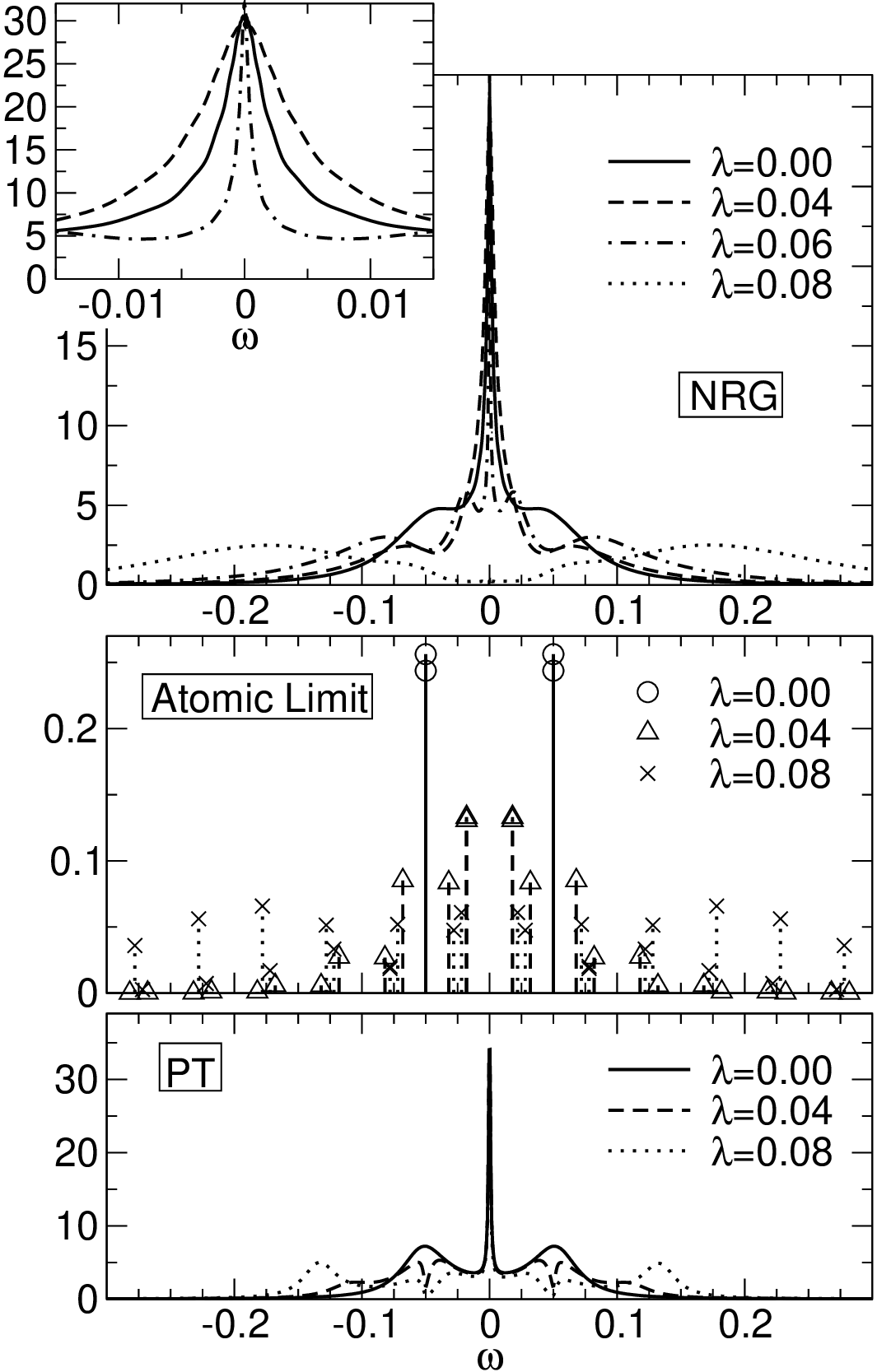}
    \caption{Spectral density
      for $U=0.1$,
      $\epsilon_f=-0.05$, $V=0.1 \rightarrow \Delta\approx 0.016$ and
      $\omega_0=0.05$. The respective values of $\lambda$ are given in
      the graph. The upper panel shows NRG results, the two lower panels
      the atomic limit and the perturbation theory (PT) results. For the
      atomic limit calculation, the height of the peaks is a measure of
      the respective spectral weight of the excitation.}
    \label{fig:dos_u01sym}
  \end{center}
\end{figure}

\begin{figure}[h]
  \begin{center}
    \includegraphics[width=0.5\textwidth]{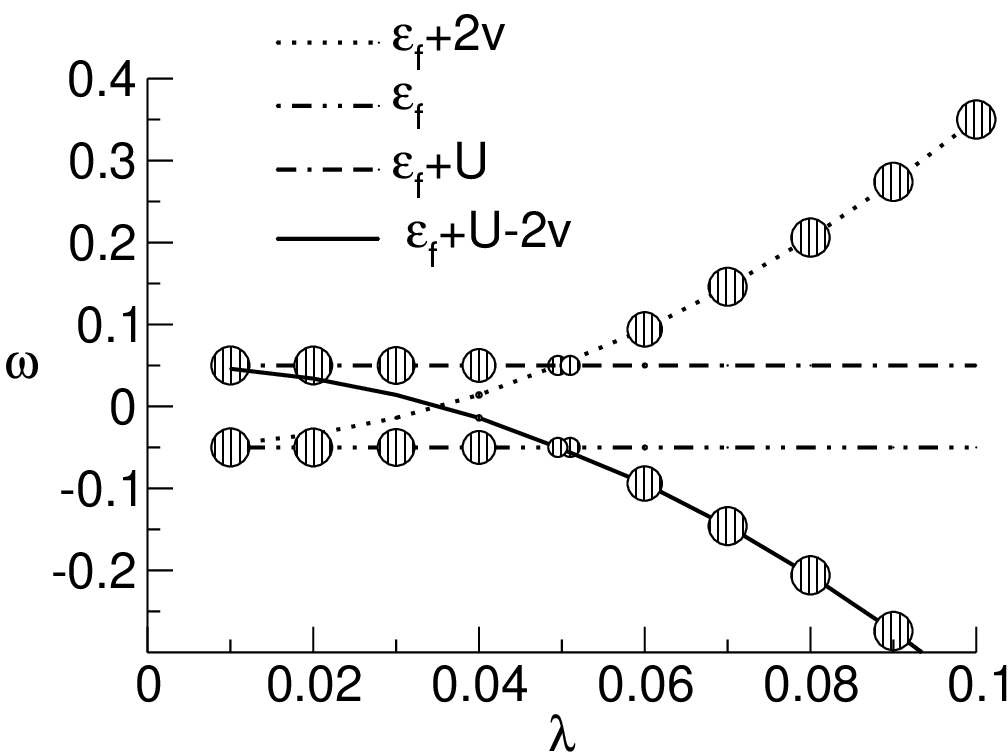}
    \caption{Positions of the dominant peaks for the atomic limit
      calculation from figure~\ref{fig:dos_u01sym}. The size of the
      circles shows the total weight for the respective type of
      excitations (see text).}
    \label{fig:pw_u01sym}
  \end{center}
\end{figure}

\begin{figure}[h]
  \begin{center}
    \includegraphics[width=0.8\textwidth]{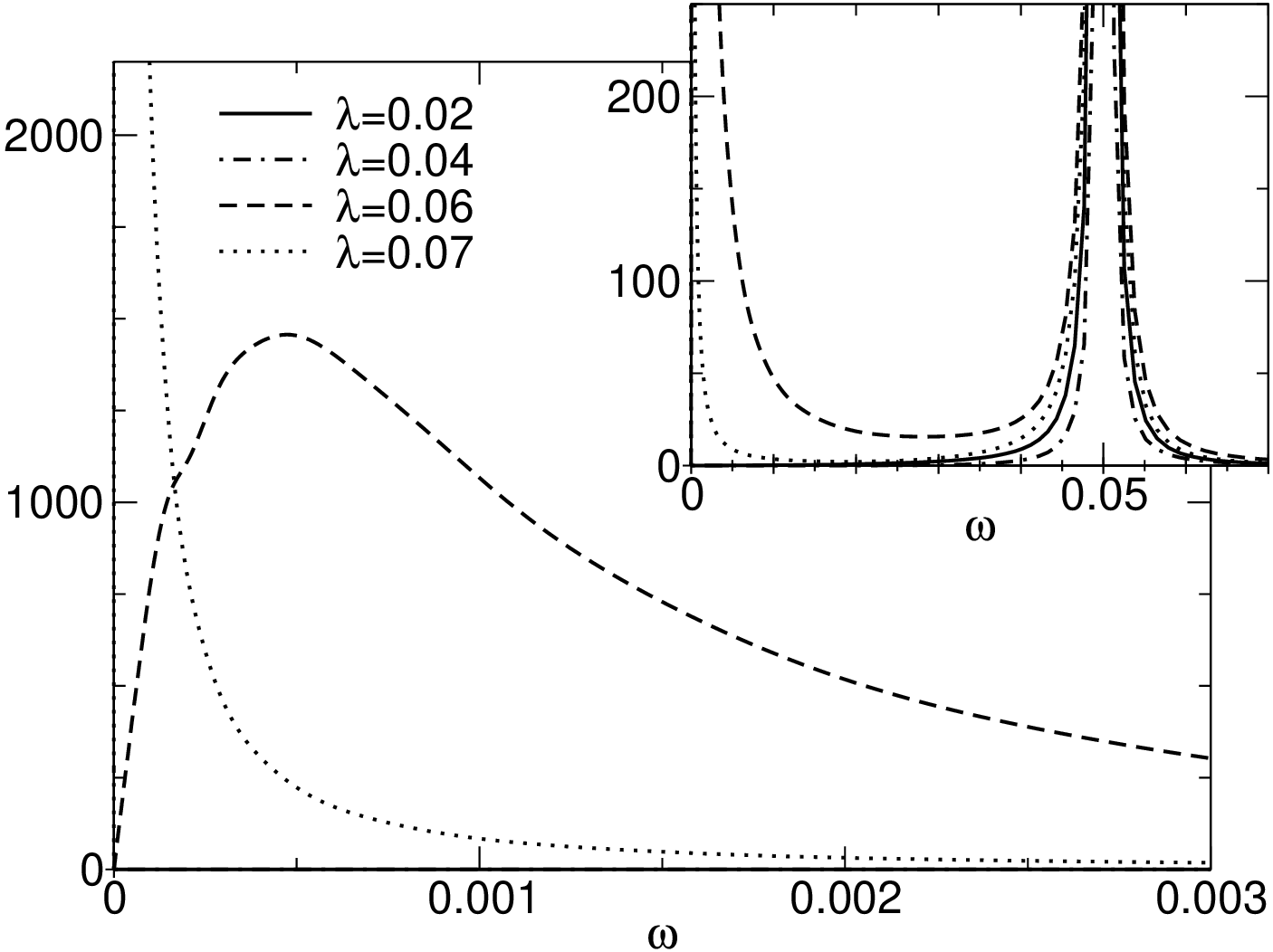}
    \caption{Imaginary part of the phonon propagator for the parameters
      of figure~\ref{fig:dos_u01sym}.}
    \label{fig:pp_u01sym}
  \end{center}
\end{figure}

In figure~\ref{fig:dos_u01sym}, we plot the spectral density for $U=0.1$
and $\epsilon_f=-0.05$. In contrast to the $U=0$ case, the central peak
initially broadens with increase of $\lambda$ and then narrows rapidly.
When $\lambda=0$, $U/\pi\Delta=2.03$, we are initially in the Kondo
regime and so we can interpret the
central peak here as Kondo resonance with a width of the order of $2 T_K
\sim (U\Delta)^{\frac{1}{2}} \, \exp(\frac{-U}{\pi\Delta})$.
Hence if $U$ is replaced by $\bar{U}= U-2v$
as in equation~(\ref{aht}) then $T_K$ would be expected to increase, provided
$\bar{U}$ remains sufficiently large so that we stay in the Kondo
regime. The situation will change as $\bar{U}$ decreases below
$\pi\Delta$. When $\bar{U}$ becomes negative we expect the
results to be similar to the $U=0$ symmetric case considered in Sec.~\ref{sec:symmetric-u=0-model}.
This is qualitatively the case, and a very rapid narrowing of the
central peak can be seen in the inset of the top panel in
figure~\ref{fig:dos_u01sym} for $\lambda>\lambda_c$ where
$\lambda_c=\sqrt{U\omega_0/2}=0.05$.

For small $\lambda$, we see the typical charge excitations of the
symmetric Anderson model at $\omega=\pm U/2$. For larger values of
$\lambda$, we see in general four high-energy peaks. The positions and
relative weights of these peaks can be understood in terms of the diagram
shown in figure~\ref{fig:pw_u01sym}, where we plot the positions of the
dominant excitation peaks taken
from the atomic limit result displayed in the middle panel of
figure~\ref{fig:dos_u01sym}. These are obtained from the atomic limit
Green's function by replacing the multi-phonon peaks for
each term in equation~(\ref{atlim}) by a single peak with $\bar{n}$
excited phonons:
\numparts
\begin{eqnarray}
  \label{excitations}
  \omega&=\bar{\epsilon_f}+\bar{n}\omega_0=\epsilon_f+2v  &\text{from zero-occupied site} \\ 
  \omega&=\bar{\epsilon_f}-\bar{n}\omega_0=\epsilon_f   &\text{from single-occupied site} \\
  \omega&=\bar{\epsilon_f}+\bar{U}+\bar{n}\omega_0=\epsilon_f+U  &\text{from single-occupied site} \\   
  \omega&=\bar{\epsilon_f}+\bar{U}-\bar{n}\omega_0=\epsilon_f+U-2v\,\,  &\text{from double-occupied site}
\end{eqnarray}  
\endnumparts
The total weight in each peak is the sum of the weights of the
contributing excitations in the atomic limit. The size of the symbols in
each curve indicates the corresponding weight factor.
One
observes the transfer of weight with increasing $\lambda$ from the
excitations in which the site is initially singly-occupied to those of the zero- and
doubly-occupied states as $\bar{U}$ changes sign. In the approach to
the regime where
these levels cross, all four peaks can be clearly distinguished in
figure~\ref{fig:dos_u01sym}.

The imaginary part of the phonon Green's function is shown in figure~\ref{fig:pp_u01sym}. We
see that while $\bar{U}>0$ charge fluctuations are suppressed and
there is no soft phonon mode $\omega_\lambda$. This mode only emerges
when $\bar{U}$ becomes negative. Its appearance clearly correlates
with the collapse of the central peak. 

For comparison, the perturbational results corresponding to the lowest-order
diagrams in $U$ and $\lambda$ are shown in the lower panel
of figure~\ref{fig:dos_u01sym}. It only gives a good approximation to the
NRG results for $\lambda=0$, and there is clear disagreement for finite
$\lambda$ as in the $U=0$ case discussed in Sec.~\ref{sec:symmetric-u=0-model}.

\section{Asymmetric $U\ne 0$ Model}
\label{sec:asymmetric-une-0}

\begin{figure}[ht]
  \begin{center}
    \includegraphics[width=0.8\textwidth]{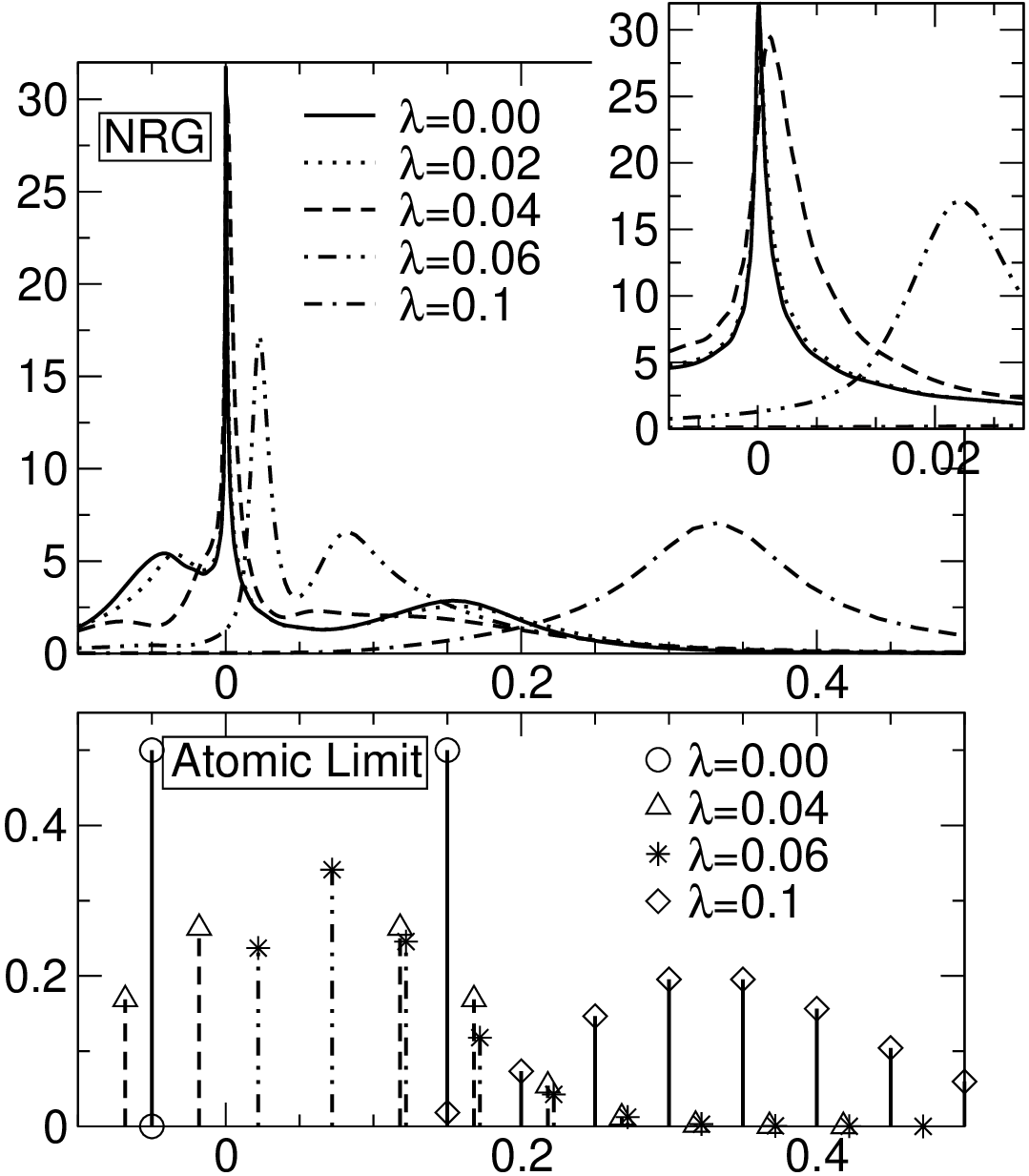}
      \caption{Same as figure~\ref{fig:dos_u01sym}, but with $U=0.2$.}
    \label{fig:dos_u02asym}
  \end{center}
\end{figure}

\begin{figure}[ht]
  \begin{center}
    \includegraphics[width=0.8\textwidth]{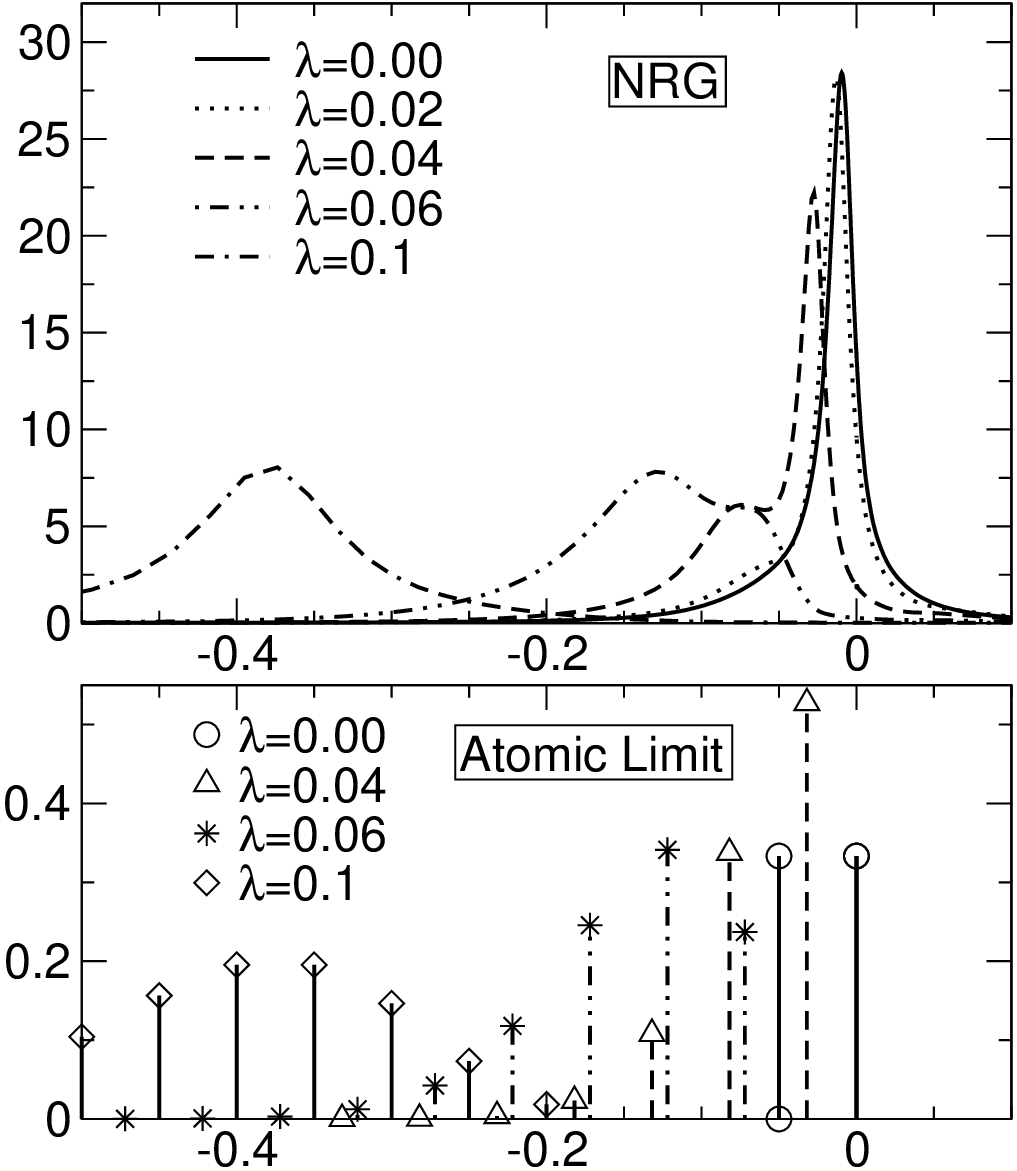}
    \caption{Same as figure~\ref{fig:dos_u01sym}, but with $U=0.05$.}
    \label{fig:dos_u005asym}
  \end{center}
\end{figure}

\begin{figure}[h]
  \begin{center}
    \includegraphics[width=0.47\textwidth]{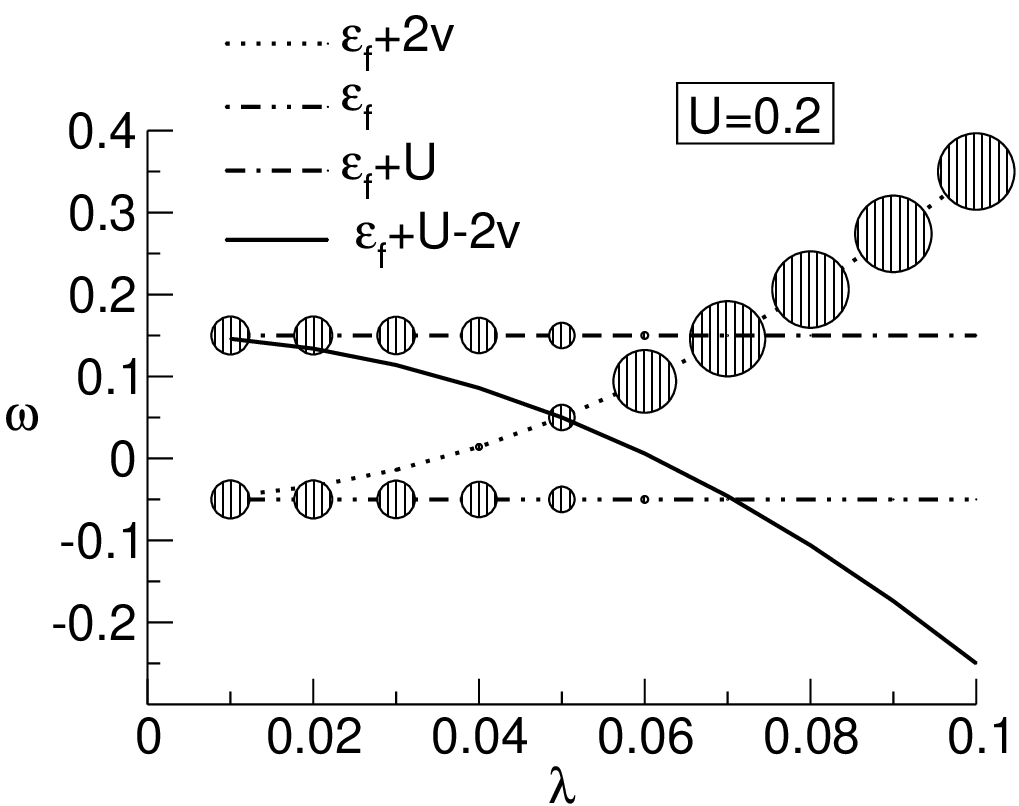}
    \includegraphics[width=0.47\textwidth]{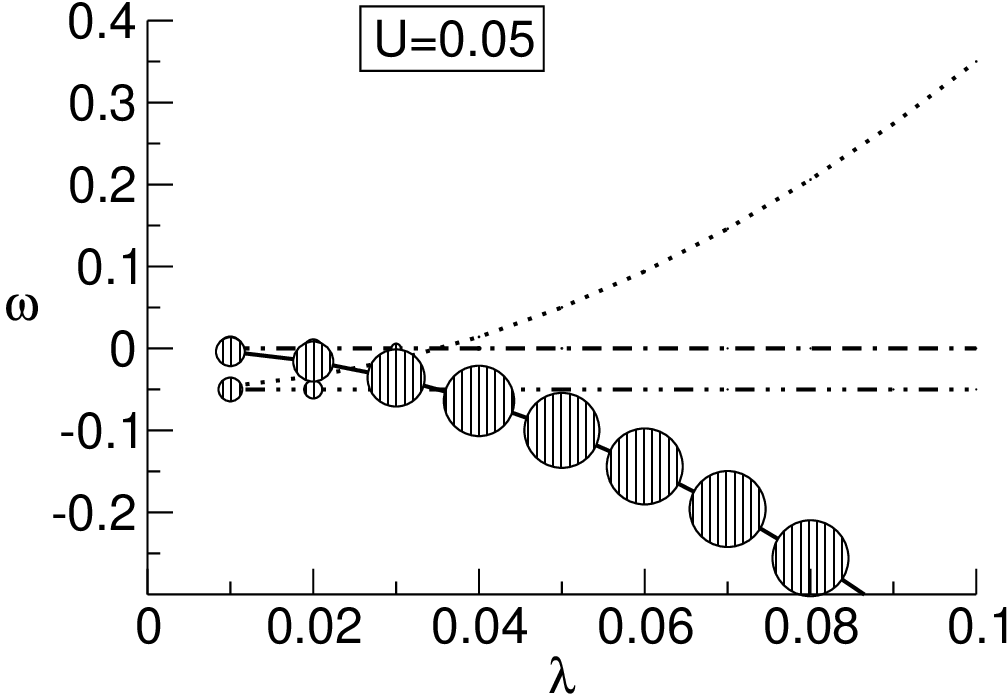}
    \caption{Positions of the dominant peaks for the atomic limit
      calculation from Figs.~\ref{fig:dos_u02asym} and
      \ref{fig:dos_u005asym}. The size of the
      circles shows the total weight for the respective type of
      excitations (see text).}
    \label{fig:pw_u02005}
  \end{center}
\end{figure}

In Figs.~\ref{fig:dos_u02asym} and~\ref{fig:dos_u005asym}, we plotted
the spectral density for $U=0.2$ and $U=0.05$, respectively
($\epsilon_f=-0.05$,$V=0.1 \rightarrow \Delta=0.016$, $\omega_0=0.05$). In the first case, for
$\lambda=0$, we are in the Kondo regime and a narrow Kondo resonance at
the Fermi level can clearly be distinguished from the atomic level
below.

For intermediate values of $\lambda$, the spectrum again involves four
dominant peaks whose relative positions and weights can be understood on
the basis of the diagram on the left in figure~\ref{fig:pw_u02005}. These excitations correspond exactly
to the ones in equation~(21). The transfer of weight with
increasing $\lambda$ is rather different: in the intermediate regime,
$0.04<\lambda<0.08$, weight is transfered from those excitations  which
involve an initial single-occupied state to those in which the initial
state is unoccupied.
As this transfer of weight develops, the Kondo resonance
disappears as can be seen clearly in the inset of the top panel of
figure~\ref{fig:dos_u02asym}. 

In the results with $U=0.05$, figure~\ref{fig:dos_u005asym} and the
diagram on the right
in figure~\ref{fig:pw_u02005}, only one
peak is to be seen for $\lambda=0$. reflecting the fact that for this
set of parameters, we are in the mixed-valence regime. The change in
high-energy features with increase of $\lambda$ can again be understood
in terms of the transfer of weights between the dominant excitations as
indicated in figure~\ref{fig:pw_u02005}. For large $\lambda$, the spectral weight
accumulates for excitations from the double-occupied local
level. Consequently, the peak moves to lower energies in a similar way
to that shown in
figure~\ref{fig:dos_u0asym} for $U=0$ (see discussion in Sec.~\ref{sec:asymmetric-u=0-model}). The
crossover occurs for smaller values of $\lambda$ compared to the case
shown in
figure~\ref{fig:dos_u02asym}
due to the smaller value of $U$.

In both these cases, the charge fluctuations are small, and there is no
evidence of a soft phonon mode developing in the phonon spectrum.
Also, perturbation theory does not give reasonable results for either
one of these cases.

\section{Conclusions}
\label{sec:conclusions}

\begin{figure}[ht]
  \begin{center}
    \includegraphics[width=0.47\textwidth]{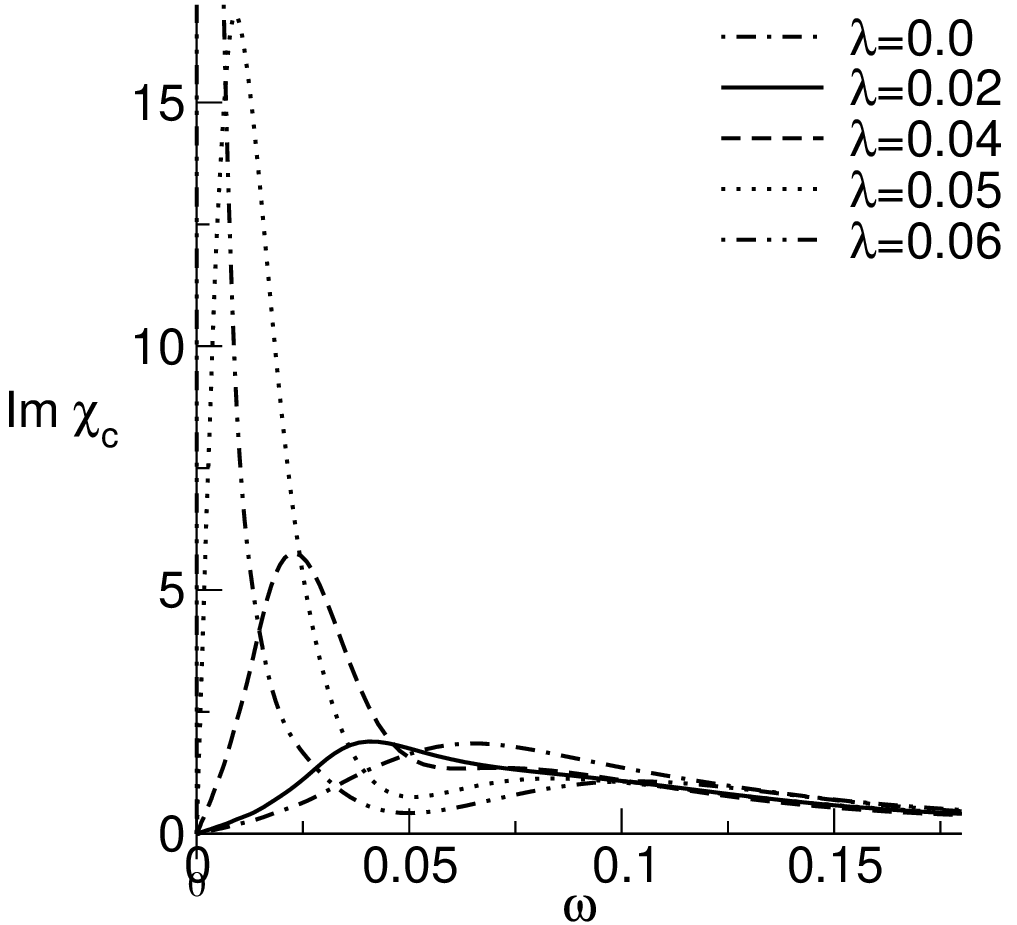}
    \includegraphics[width=0.47\textwidth]{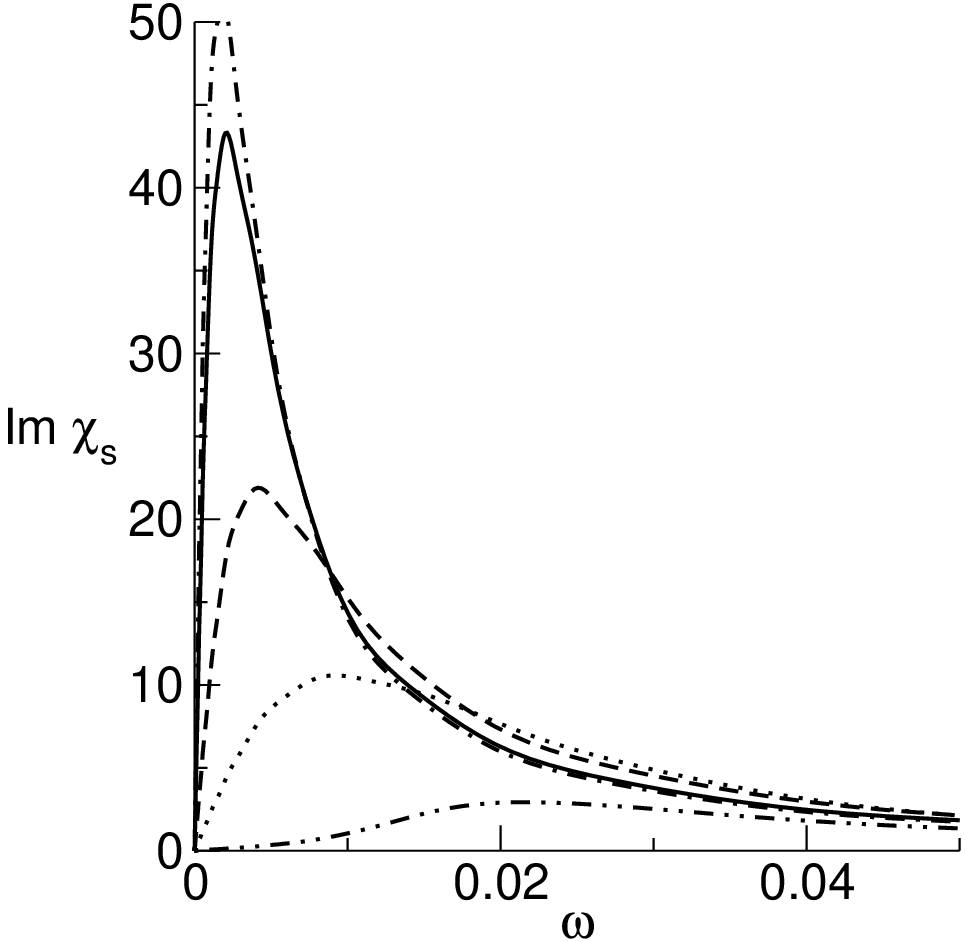}
    \caption{Imaginary part of the dynamic charge ($\chi_c$) and spin
      susceptibilities ($\chi_s$) for the symmetric
      parameters of figure~\ref{fig:dos_u01sym} ($U=0.1$,
      $\epsilon_f=-0.05$, $V=0.1$). }
    \label{fig:sus_u01}
  \end{center}
\end{figure}

We begin with an overview of the detailed results presented in sections 3-6,
and look first of all at the results for the symmetric model. When  $\lambda=0$
and we move from $U=0$ to $U=0.1$, we move from an intermediate valence regime,
in which the spin and charge fluctuations are of the same order to the Kondo regime,
in which the charge fluctuations are suppressed. The central resonance narrows
considerably from a width $2\Delta$ to one determined by the Kondo temperature $T_{\rm K}$,
and atomic-like high energy peaks begin to develop. When we switch on, and increase,
the phonon coupling the central peak initially broadens and then for large phonon
coupling strength narrows markedly. The initial broadening can be understood as mainly due
to the reduction in the effective electron interaction from $U\to\bar U=U-2v$, causing
an increase in $T_{\rm K}$, which depends exponentially on $U$. However, the atomic-like
peaks at $\omega\sim \pm U/2$, seem to be largely unaffected. As $\bar U$ falls below
the value $\pi\Delta$, the relative strength of the charge fluctuations increases, and
we have a regime which is similar to the intermediate valence regime as the spin and
charge fluctuations are of the same order, though high energy features associated with the
atomic limit persist. Finally we move into the regime where $\bar U$ reaches zero and then
becomes negative. In this limit we see that the spin fluctuations are largely
suppressed as pairing occurs and the impurity state is either doubly-occupied or
empty. Excitations from these states are associated with broadened high
energy features above and below the Fermi-level.
The central resonance narrows exponentially. 
One can interpret the narrow resonance in the strong-coupling regime as
analogue of the Kondo resonance in a negative-$U$ model. In this regime
the spin fluctuations have been largely eliminated but there remain long-lived
charge fluctuations between the double and zero occupation states at the impurity site.
The double and zero occupied states can be interpreted as the
up and down configurations of a local isospin. In the $\sigma-\tau$ model
an exchange coupling of such an isospin to the conduction electrons is invoked
to mimic one of the channels in the two channel Kondo model~\cite{CIT95,BBHZ99}.

The reversal of
roles of the spin and charge fluctuations can be seen in the results for the corresponding
dynamic susceptibilities shown in figure~\ref{fig:sus_u01} for
$U=0.1$. For $\lambda=0$ there is a low energy peak in the
spin susceptibility that gradually gets suppressed for increasing $\lambda$ and moves to
higher frequencies. The reverse trend is seen in the charge susceptibility where a low
energy peak develops and becomes very narrow in the negative $\bar U$
range.

We can also interpret the narrow central resonance as the zero-phonon
excitation which carries very little weight in the strong-coupling
regime because of the high probability of exciting a phonon when
changing the impurity occupation. The narrow peak in the charge
susceptibility is reflected in the phonon spectrum as low energy mode at
$\omega_\lambda$. 

Whereas the first of these interpretations would give a resonance that
exponentially narrows as $e^{-|\bar{U}|/8\pi\Delta}$ with
$\bar{U}-U\sim -\lambda^2/\omega_0$, we found that it narrows more rapidly than
this expression would indicate (see figure~\ref{fig:peakwidth}).
The expression for the narrowing involving the soft phonon mode,
$\sim \exp(-v/\omega_\lambda)$ with $\omega_\lambda \sim
\exp(-\lambda^2/\omega_0^2)$ gives a better fit to data.

In the results for the asymmetric model the low energy phonon mode does not develop, and there
is no narrow resonance at the Fermi-level other than the original Kondo peak  in the
case $U=0.2$ at zero or
small phonon coupling. In the large $\lambda$ regime ($\bar U<0$) in the two cases studied
there is either a broad
peak below the Fermi-level corresponding to double occupation, or one above the Fermi-level
corresponding to zero occupation. For asymmetric conditions the double and zero occupied
states are not equally likely, and the local level position $\epsilon_f$
for $\epsilon_f\ne 0$ acts like a magnetic field
on the isospin suppressing the `Kondo' resonance. The same behaviour is
observed in the negative-$U$ Anderson model without phonons.

Our conclusions are very much in line with those of Hewson and Newns in their
study of the spinless model~\cite{HN79,HN80}. A direct comparison with their work is not possible
because no $\bar U$-term can be induced by the phonons in the spinless
case. 
Their conclusion was that a significant narrowing of the local resonance
could only be induced by strong phonon coupling if the impurity level lies at,
or very close to the Fermi-level (intermediate valence regime),
when the zero or single occupation states of the local level have the same energy,
This closely resembles the situation in the model with spin, but in
terms of
doubly-occupied and the empty states.

We have surveyed the physically interesting
parameter regimes of the model using the NRG, and related our results
to atomic limit calculations and perturbation theory. The relevant parameter regime for magnetic
impurities systems is expected to be large $U$, with a phonon
coupling such $v\ll U$, and $\omega_0<\Delta$. We have taken
a relatively large value of $\omega_0$ ($\omega_0\sim\pi\Delta$) in order
to see the features in the spectrum clearly and save computational time
(see Appendix B), but results for smaller
values of $\omega_0$ were found to be qualitatively similar.

Dynamical response functions for impurity systems are difficult to
observe experimentally, so it would be interesting
to calculate the effects of the phonon coupling on the thermodynamic
behaviour which could be more readily compared with experiment.
This we propose to do for a separate publication. As a more realistic
model for magnetic impurities we also intend to include a direct coupling
of the phonon mode to the hybridization term, because this term should be
rather sensitive to local lattice vibrations.

Having taken the first step of calculating the dynamic properties of the
impurity model with phonons, we now propose to extend these calculations
to lattice models within the framework of the dynamical mean-field theory.
These calculations should be particularly relevant to high-$T_c$
superconductors where recent photoemission experiments give evidence of
strong electron-phonon coupling~\cite{Lea01}. The electron-lattice
coupling is also known to be important for the manganites~\cite{Ram97}. The
calculation could be extended to a model for these systems by including a coupling to a
local spin~\cite{MLS95}, which would require only a generalization of the
impurity part of the Hamiltonian. The calculation could also be extended to
include a dispersive phonon mode. This would require the introduction of a second chain
of states, but then only a lower proportion of states could be kept in each iteration.

\ack
We wish to thank 
to the  EPSRC  for the support of a research grant (GR/J85349), and the Newton Institute, Cambridge,
where this work was initiated during their six-month programme on `Strongly Correlated Electron Systems'. 
We also thank Simon Bradley for his work in the development stage of the
NRG program and Ralf Bulla for helpful discussions.

\appendix

\section{Equations of motion}
We can use the standard equation of motion for the Fourier transform of the
double-time Green's function,
\begin{equation}
  \langle\langle A,B
  \rangle\rangle_{\omega}=\langle [A,B]_\eta \rangle+\langle\langle [A,H]_{-},B
  \rangle\rangle_{\omega}
\end{equation}
To derive the expression for the self-energy given in equation~(\ref{eq:sigma}),
we choose $A=f^{}_\sigma$ and $B=f^{\dagger}_\sigma$
and $\eta=+$ as these are fermion operators.
The equations of motion give
\begin{equation}
  \fl (\omega-\epsilon_f)G_\sigma(\omega)=1+U\langle\langle f^{}_\sigma f^{\dagger}_{\bar\sigma} 
  f^{}_{\bar\sigma} ,f^{\dagger}_\sigma\rangle\rangle_{\omega}+\lambda \langle\langle f^{}_\sigma
  (b^{\dagger}+b) ,f^{\dagger}_\sigma\rangle\rangle_{\omega}
  +\sum_{\bf k}V_{\bf k}\langle\langle c^{}_{{\bf k}\sigma}
  ,f^{\dagger}_\sigma \rangle\rangle_{\omega}
\end{equation}
\begin{equation}
  (\omega-\epsilon_{\bf k})\langle\langle c^{}_{{\bf k}\sigma} ,f^{\dagger}_\sigma \rangle\rangle_{\omega}
  =V_{\bf k}G_{\sigma}(\omega)
\end{equation}
Hence 
\begin{equation}(\omega-\epsilon_f -\Delta(\omega))G_\sigma(\omega)=1+UF_{\sigma}(\omega)+\lambda M_\sigma(\omega)\end{equation}
where 
\begin{equation}F_{\sigma}(\omega)=\langle\langle f^{}_\sigma f^{\dagger}_{\bar\sigma} 
f^{}_{\bar\sigma} ,f^{\dagger}_\sigma
\rangle\rangle_{\omega}  ,\quad M_\sigma(\omega)=
\langle\langle f^{}_\sigma (b^{\dagger}+b) ,f^{\dagger}_\sigma\rangle\rangle_{\omega}\end{equation}
We define
\begin{equation}
\Sigma_\sigma^{U}(\omega)=U{F_\sigma(\omega)\over G_\sigma (\omega)},\quad
\Sigma_\sigma^{\lambda}(\omega)=\lambda{M_\sigma(\omega)\over G_\sigma (\omega)}\end{equation}
so that $\Sigma_\sigma(\omega)=\Sigma_\sigma^{U}(\omega)+\Sigma_\sigma^{\lambda}(\omega)$.\par
\smallskip
For the boson Green's function $D(\omega)$  we take $A=B=b^{\dagger}+ b$ and $\eta=-$.
The equation of motions of motion give
\begin{equation}
  \omega D(\omega)=
  \omega_0\langle\langle(-b^{\dagger}+ b),( b^{\dagger}+
  b)\rangle\rangle_{\omega}
\end{equation}
\begin{equation}
  \omega\langle\langle(-b^{\dagger}+ b),( b^{\dagger}+ b)\rangle\rangle_{\omega}=2+2\lambda
  \langle\langle\hat Q,( b^{\dagger}+
  b\rangle\rangle_{\omega}+\omega_0D(\omega)
\end{equation}
Hence,
\begin{equation}
  D(\omega)=D^0(\omega)+\lambda D^0(\omega)\langle\langle\hat O,(
  b^{\dagger}+ b)\rangle\rangle_{\omega}
\end{equation}
where $\hat O=\sum_{\sigma}f^{\dagger}_\sigma f^{}_\sigma -1$.
Taking the equation of motion for the right-hand operator,
\begin{equation}
  -\omega\langle\langle\hat O,(b^{\dagger}+ b)\rangle)\rangle_{\omega}=
  \omega_0\langle\langle \hat O,(-b^{\dagger}+ b)\rangle\rangle_{\omega}
\end{equation}
\begin{equation}
  -\omega\langle\langle\hat O,(-b^{\dagger}+ b)\rangle\rangle_{\omega}=2\lambda
\langle\langle\hat O,\hat O\rangle\rangle_{\omega}+
\omega_0\langle\langle\hat O,( b^{\dagger}+ b)\rangle\rangle_{\omega}
\end{equation}
\begin{equation}
  \langle\langle \hat O,(b^{\dagger}+ b\rangle)\rangle_{\omega}=D^0(\omega)+\lambda D^0(\omega)
\langle\langle \hat O,\hat O\rangle\rangle_{\omega}
\end{equation}
Hence the result,
\begin{equation}  
  D(\omega)=D^0(\omega)+\lambda^2D^0(\omega)\langle\langle \hat O,\hat
  O\rangle\rangle_{\omega}D^0(\omega)
\end{equation}

\section{Matrix elements and NRG calculation}
To evaluate the Green's function $M_{\sigma}(\omega)$, and hence the self-energy $\Sigma_\sigma^{\lambda}(\omega)$, we need to calculate the matrix elements of the
operator $f^{}_\sigma(b^{\dagger}+b)$. We can classify the states according to total charge and total spin quantum numbers $Q$,$S$ and $S_z$, and an additional index $w$
or $r$, and we define reduced matrix elements using the Wigner-Eckart theorem in the usual way.
 The bose operators only enter explicitly in the
calculation at the impurity site. Let the basis states used for this calculation be denoted
by $|Q,S,S_z,n,r\rangle$ where $n$ denotes the occupation number for the bose states, and $r$ is an additional index. After diagonalization
the exact eigenstates at the impurity site can be written in the form $|Q,S,S_z,w\rangle$,
where $w$ no longer denotes the number of bosons, which are not conserved by the interaction term,
so $w$ is a general index.
We need the reduced matrix elements between these eigenstates, $\langle Q,S,w||f^{}_\sigma(b^{\dagger}+b)||Q',S',w'\rangle$.
If 
\begin{equation}
  |Q',S',w'\rangle=\sum_{n'} U_{Q',S'}(w',n'r')|Q',S',n',r'\rangle 
\end{equation}
then
\begin{equation}
  \eqalign{
    \fl  \langle Q,S,w||f^{}_\sigma(b^{\dagger}+b)||Q',S',w'\rangle=\\
    =\sum_{n,n',r,r'}U_{Q,S}(w,n,r)U_{Q',S'}(w',n'r') \langle Q,S,n,r||f^{}_\sigma(b^{\dagger}+b)
    ||Q',S',n',r'\rangle\\
    \eqalign{=\sum_{n,r,r'} U_{Q,S}(w,n,r) (\sqrt{n}U_{Q',S'}(w',(n-1)r')+\\
      +\sqrt{n+1}U_{Q',S'}(w',(n+1)r')) \langle Q,S||f^{}_\sigma 
      ||Q',S'\rangle 
  }}
\end{equation}
as $\langle Q,S,n,r||f^{}_\sigma
||Q',S',n,r'\rangle$ is independent of $n$. Hence we can deduce the  values of 
$\langle Q,S,w||f^{}_\sigma(b^{\dagger}+b)||Q',S',w'\rangle$ after diagonalization at the impurity
site. Once we have these initial values the equations at all later steps will take the same form as
those for  $\langle Q,S,w||f^{}_\sigma||Q',S',w'\rangle$.\par
The spectral density $C_\sigma(\omega)$ of the Green's function $M_\sigma (\omega)$
is given by 
\begin{equation}
  \fl C_\sigma(\omega)={1\over Z}\sum_{n,n'}\langle
  n|(b+b^{\dagger})f^{}_\sigma | n'\rangle\langle n' | 
  f^{\dagger}_\sigma | n\rangle \delta(\omega-(E_{n'}-E_n))(e^{-\beta E_n}
  +e^{-\beta E_{n'}} ) 
\end{equation} 
where $|n\rangle=|Q_n,S_n,S_{z,n},w_n\rangle$ and $Z$ is the partition function. When expressed in terms 
of reduced matrix elements it takes exactly the same form as the spectral density for $G_\sigma$
except that the reduced matrix element  $\langle Q,S,w||f^{}_\sigma(b^{\dagger}+b)||Q',S',w'\rangle$
replaces  $\langle Q,S,w||f^{}_\sigma||Q',S',w'\rangle$.\\[0.3cm]

\noindent\textbf{NRG parameters:}\\
The NRG calculations were performed using the discretization parameter
$1.7<\Lambda<2.$, and keeping at least 600 states in each iteration (see
Ref.~\cite{KWW80a,BCV01}). The number of eigenstates of an impurity site coupled to
a local phonon mode (the atomic limit~(\ref{atlim})), is already
infinite. So a cutoff of higher-energy states is already necessary at
the initial step of the 
NRG procedure. The probability distribution function to find
$n$ excited phonons in the system has a clear maximum around
$n=\bar{n}=v/\omega_0$, as can be read off
equation~(\ref{atlim}), and falls off rapidly for larger
values of $n$.
So in the spirit of the NRG it is sufficient to keep only
states with $n<n_{\rm cutoff}$ where we chose an initial $n_{\rm cutoff}=4\bar{n}$.

\section{Derivation of equations~(\ref{pf2}) and~(\ref{pf3})}

Here we give details of the explicit removal of the phonon terms in the Hamiltonian to derive
equations (\ref{pf2}) and (\ref{pf3}). We will factorise the time-ordering operator $T$ in equation (\ref{pf}) so that $T=T_{\rm F}T_{\rm B}$,
where $T_{\rm F}$ operates purely on the Fermi operators and $T_{\rm B}$ purely on the bose ones.
As the interaction term with the phonons,
\begin{equation}  
  H^{\rm B}_{\rm int}(\tau)=\lambda\hat
  O(\tau)(b^{\dagger}e^{\omega_0\tau}+be^{-\omega_0\tau}),
\end{equation}  
is linear in $b$ and $b^{\dagger}$, we can apply the well-known
Baker-Hausdorff formula to deduce
\begin{equation}
  T_{\rm F}T_{\rm B}e^{-\int_0^\beta H_{\rm int}(\tau)d\tau}=T_{\rm F}e^{-\int_0^\beta H_{\rm int}(\tau)
    d\tau}
  e^{{1\over 2}\int_0^\beta\int_0^\tau [H_{\rm int}(\tau),H_{\rm
      int}(\tau')]d\tau d\tau'}.
\end{equation}  
We now take the expectation value of
$e^{-\int_0^\beta H_{\rm int}(\tau) d\tau}$
with respect to the boson states using $\langle e^A \rangle=e^{{1\over 2} \langle A^2 \rangle}$,
applicable when $A$ is a linear combination of $b$ and $b^{\dagger}$~\cite{Mahbook}.
We find in the exponential,
\begin{equation}  
  {\lambda^2\over 2}\int_0^\beta\int_0^\tau \hat O(\tau)\hat O(\tau')(\gamma(\tau-\tau')
  +\gamma(\tau'-\tau))d\tau d\tau' -\int_0^\beta H^{\rm F}_{\rm
  int}(\tau)d\tau, 
\end{equation}  
where $H^{\rm F}_{\rm int}(\tau)$ is the purely fermion interaction term, and
\begin{equation}  \gamma(\tau-\tau')=n(\omega_0)e^{\omega_0(\tau-\tau')}+(1+n(\omega_0))e^{-\omega_0(\tau-\tau')}.\end{equation}  
We also have that
\begin{equation}  [H_{\rm int}(\tau),H_{\rm int}(\tau')]
={\lambda^2\over 2}\hat O(\tau)\hat O(\tau')(e^{-\omega_0(\tau-\tau')}-e^{\omega_0(\tau-\tau')}),\end{equation}  
subject to the fermion time-ordering within the complete expression.\par
Finally, we find
\begin{equation}  
  Z/Z_0=\langle T_{\rm F}e^{-\int_0^\beta H^{\rm F}_{\rm
      int}(\tau)d\tau}e^{-{\lambda^2}\int_0^\beta\int_0^\tau \hat
    O(\tau)\hat O(\tau')D^0(\tau-\tau')d\tau d\tau'}\rangle_{\rm F},
\end{equation}  
where 
\begin{equation}  
  D^0(\tau-\tau')=\left(
    {{e^{-\omega_0|\tau-\tau'|}}\over{1-e^{-\omega_0\beta}}}+
    {{e^{\omega_0|\tau-\tau'|}}\over{e^{\omega_0\beta}-1}}\right),
\end{equation}  
which is the imaginary time representation of the phonon Green's function.\par



\vspace{1cm} 

\textbf{References}

\begin{thebibliography}{10}

\bibitem{Hol59}
T.~Holstein,
 Ann. Phys. {\bf 8}, 325 1959.

\bibitem{GKKR96}
A.~Georges, G.~Kotliar, W.~Krauth, and M.~J. Rozenberg,
 Rev. Mod. Phys. {\bf 68}(1), 13 1996.

\bibitem{MV89}
W.~Metzner and D.~Vollhardt,
 Phys. Rev. Lett. {\bf 62}, 324 1989.

\bibitem{Wil75}
K.~Wilson,
 Rev. Mod. Phys. {\bf 47}(4), 773 1975.

\bibitem{KWW80a}
H.~R. Krishna-murthy, J.~W. Wilkins, and K.~G. Wilson,
 Phys. Rev. B {\bf 21}(3), 1003 1980.

\bibitem{KWW80b}
H.~R. Krishna-murthy, J.~W. Wilkins, and K.~G. Wilson,
 Phys. Rev. B {\bf 21}(3), 1044 1980.

\bibitem{SSK89}
O.~Sakai, Y.~Shimizu, and T.~Kasuya,
 J. Phys. Soc. Japan {\bf 58}(10), 3666 1989.

\bibitem{CHZ94}
T.~A. Costi, A.~Hewson, and V.~Zlatic,
 J. Phys.: Condens. Matter {\bf 6}, 2519 1994.

\bibitem{SK94}
O.~Sakai and Y.~Kuramoto,
 Solid State Commun. {\bf 89}, 307 1994.

\bibitem{Bul99}
R.~Bulla,
 Phys. Rev. Lett. {\bf 83}, 136 1999.

\bibitem{BCV01}
R.~Bulla, T.~Costi, and D.~Vollhardt,
 Phys. Rev. B {\bf 64}, 045103 2001.

\bibitem{And61}
P.~W. Anderson,
 Phys. Rev. {\bf 124}(1), 41 1961.

\bibitem{HN79}
A.~C. Hewson and D.~M. Newns,
 J. Phys. C {\bf 12}, 1665 1979.

\bibitem{HN80}
A.~C. Hewson and D.~M. Newns,
 J. Phys. C {\bf 13}, 4477 1980.

\bibitem{BHP98}
R.~Bulla, A.~C. Hewson, and T.~Pruschke,
 J. Phys.: Condens. Matter {\bf 10}, 8365 1998.

\bibitem{NObook}
J.~W. Negele and H.~Orland,
 {\em Quantum Many-Particle systems},
 Addison-Wesley 1988.

\bibitem{Mahbook}
G.~D. Mahan,
 {\em Many-Particle Physics},
 Plenum Press 1990.

\bibitem{CIT95}
P.~Coleman, L.~Ioffe, and A.~M. Tsvelik,
 Phys. Rev. B {\bf 52}, 6611 1995.

\bibitem{BBHZ99}
S.~C. Bradley, R.~Bulla, A.~C. Hewson, and G.-M. Zhang,
 Eur. Phys. J. B {\bf 11}, 535 1999.

\bibitem{Lea01}
A.~Lanzara, P.~V. Bogdanov, X.~J. Zhou, S.~A. Keller, D.~L. Feng, E.~D. Lu,
  T.~Yoshida, H.~Eisaki, A.~Fujimori, K.~Kishio, J.-I. Shimoyama, T.~Nodaand~S.
  Uchida, Z.~Hussain, and Z.-X. Shen,
 Nature {\bf 412}, 510 2001.

\bibitem{Ram97}
A.~P. Ramirez,
 J. Phys.: Condens. Matter {\bf 9}, 8171 1997.

\bibitem{MLS95}
A.~J. Millis, P.~B. Littlewood, and B.~I. Shraiman,
 Phys. Rev. Lett. {\bf 74}(25), 5144 1995.

\end{thebibliography}
\end{document}